\documentclass[12pt,preprint]{aastex}

\def\eff{{\rm eff}}

\def\e{{\rm E}}

\begin{document}
\title{Generalized Microlensing Effective Timescale}

\author{Andrew Gould}
\affil{Department of Astronomy, Ohio State University,
140 W.\ 18th Ave., Columbus, OH 43210, USA; 
gould@astronomy.ohio-state.edu}

\begin{abstract}
The microlensing effective timescale $t_\eff=\beta t_\e$ is used frequently
in high-magnification ($\beta\ll 1$) microlensing events, 
because it is better constrained than either the impact parameter $\beta$ 
or the Einstein timescale $t_\e$ separately.  It also facilitates intuitive 
understanding of lightcurves prior to determination of a model.  
Similar considerations may apply to very low magnification
events. I therefore provide a generalization of this quantity to all
events:
$$
t_\eff = \beta t_\e\sqrt{(1+\beta^2/2)(1+\beta^2/4)}.
$$

\end{abstract}


The standard point-lens/point-source microlensing magnification curve
\citep{einstein36,pac86}
\begin{equation}
A[u(t)] = {u^2 + 2\over u\sqrt{u^2+4}};
\qquad
u[\tau(t)] = \sqrt{\tau^2 +\beta^2};
\qquad
\tau(t) = {t-t_0\over t_\e}
\label{eqn:pac}
\end{equation}
is frequently rewritten in terms of the effective timescale $t_\eff$,
\begin{equation}
u(t) = \beta\sqrt{1 + {(t-t_0)^2\over t_\eff^2}},
\qquad
t_\eff \equiv \beta t_\e,
\label{eqn:teff}
\end{equation}
where $t_0$ is the time of peak, $\beta$ is the impact parameter (normalized
to the Einstein radius), and $t_\e$ is the Einstein crossing time.
This is because, particularly for high-magnification events ($\beta\ll 1$),
$t_\eff$ is often well determined from the peak of the
lightcurve alone, while $\beta$ and $t_\e$ are either highly correlated
(e.g., \citealt{mb11293}) or almost unconstrained as separate parameters 
\citep{gould96}.
This parameterization remains useful in planetary microlensing events
because the structure of the peak is often similar to that of point-lens
events.

To date, there has been relatively little work on microlensing events
at the opposite extreme, $\beta\ga 1$, in part because they are less
sensitive to planets \citep{gould92,griest98}.  However, the discovery
of a huge population of planets that are either free-floating or at
least relatively far from their hosts \citep{sumi11} implies that
the search for putative hosts (necessarily at high $\beta$) will become
more important in the future.  Indeed, there is already a planet 
(MOA-bin-1) detected in an event with $\beta\sim 1.6$ \citep{moabin1}.
Many more can be expected in the future as ``second generation'' microlensing
surveys continuously monitor all events, not just those most immediately
sensitive to planets.

I therefore propose a generalization of the effective timescale that
is appropriate for all impact parameters:
\begin{equation}
t_{\rm eff} = \biggl(-{d^2\ln A\over dt^2}\bigg|_{t=t_0}\biggr)^{-1/2}
= \beta t_{\rm E} \sqrt{(1+\beta^2/2)(1+\beta^2/4)}.
\end{equation}
In words, $t_{\rm eff}$ is the inverse square root of the second logarithmic
derivative of the magnification with respect to time.  Note that this
definition reduces to the standard formula for $\beta\ll 1$
(Eq.~(\ref{eqn:teff})).

If $t_{\rm eff}$ is a fitting variable in, e.g., a Markov Chain, then
it is essential to be able to invert this formula in order to derive
$\beta$, which is used in the actual lightcurve calculations,
and is in general a signed quantity \citep{gould04}.  I find
\begin{equation}
\beta = \sqrt{2(y-1)}{t_\eff\over |t_\eff|},
\label{eqn:invert}
\end{equation}
where
\begin{equation}
y = \sqrt{4\over 3}\cosh{\phi\over 3};
\qquad
\phi = \cosh^{-1}x
\qquad
(x>1),
\label{eqn:dgzero}
\end{equation}
\begin{equation}
y = \sqrt{4\over 3}\cos{\phi\over 3};
\qquad
\phi = \cos^{-1}x
\qquad
(x<1),
\label{eqn:dlzero}
\end{equation}
and
\begin{equation}
x \equiv \sqrt{27\over 4}\biggl({t_\eff\over t_\e}\biggr)^2.
\label{eqn:deltadef}
\end{equation}
\acknowledgments

This work was supported by NSF grant AST 1103471.

\end{document}